\documentstyle[11pt]{article}
\begin{document}
\begin{center}
{\Large Another Dash into the Record Books (wind: -2.1 m/s)} \\
\vspace{5mm}
J. R. Mureika \\
{\it Department of Computer Science \\
University of Southern California \\
Los Angeles, CA~~90089-2520}
\end{center}

When I first performed some of the analysis herein, I thought: ``Wow!  Bailey
gets another title!''.  Unfortunately, that was before August 3rd, and the 
100m final in Athens.  Now, Donovan Bailey doesn't hold the World sprint crown.
It's slipped and fallen south of the border (your border, that is!).  Maurice 
Greene, a relative unknown with a PB of 10.43s three years ago, reigned 
victorious at 9.86s, matching the third best legal clocking ever.  But, 
the events of that day didn't change the overall result of my work.
By virtue of the race in question, some would say that Bailey is no
longer the ``World's Fastest Man''.  But has he lost his place in the record
books altogether?  Currently, the 9.84s from Atlanta is still holding strong.
Can't we say more about our current international sprint sensation?

The truth is: not only can we say more about Bailey (who still seems to
have the knack of saying too much about his competition!), but we can say
something great about the ``Blast off! guy'' himself (to quote Bogdan
Poprawski), Robert Esmie, perhaps our sprint celebrity for Syndey in 2000?

Following the National Championships in Abbotsford, the media jumped all over
Bailey's winning performance.  The World Record holder had run a mediocre,
sub-par 10.03s, well off his 9.84 WR performance of 1996.  Even a medal hope
looked grim for the defending champion, who was himself displeased with his 
race, moaning and groaning of leg injuries and viruses.  In defense of his
prodigy, Texas-based coach Dan Pfaff was quick to lash back at the media, 
highlighting the underplayed fact that Bailey's run was into a head-wind, 
and a strong one at that: -2.1 m/s! He denounced the ongoing criticism of the 
time, and ventured to guess that it would probably correspond to about 9.90s 
under ideal conditions.

Taking a note from Pfaff, I quickly crunched this data in my wind-correction
formula, which was presented in the July 1997 issue of Athletics \cite{ath1}.
10.03s (how mediocre!), and -2.1 m/s...  Suffice it to say, Dan Pfaff was
wrong.  Bailey hadn't run a 9.90, nor had he even clocked anything in the
9.9 range.  In fact, Bailey had cruised to a 9.89s still-air equivalent!  
At the time, this ranked him as the world leader on my wind-correction 
tables, numerically matching the {\it legal} world leader, Ato Boldon's early
season 9.89s from Modesto, CA (run with a +0.8 m/s tailwind, a 9.94s 
still-air sprint; see Table~\ref{table5}).\\

\noindent {\bf Yet another home-town boy does good!} \\

If Bailey's head-wind charge of 10.03s was actually a blistering 9.89s, what
then became of the other times?  Table~\ref{table2} tells all: Robert
Esmie (unofficially) joined the elite club of sub-10s sprinters, with
his 10.10s in Abbotsford translating to a 9.96s still-air dash.  Bruny
Surin was knocking at the door of the club with a 10.01s equivalent, 
but couldn't quite get in this year.  Not only was Esmie's silver medal
performance good enough to put him unofficially under 10s, it also ranks
him (as of the World Championships) sixth best in the world,
behind Tim Montgomery, Frank Fredericks, Boldon, Bailey, and Greene 
(see Table~\ref{table3})!

What's interesting about that top 6 group of athletes?  Several things.
Two are from Canada, and two are from the US -- Canada ties the US for
number of top 6 100m sprinters!  What else is intersting?
Robert Esmie *doesn't* train or live in the United States, a fact shared
by the remaining athletes (Fredericks earned a CS degree
from BYU in Provo, Utah) which was quickly pointed out in NBC's coverage
of the World Championships.  Blast off indeed -- chalk one up for the
Canadians!\\

\noindent {\bf Alas, poor Bailey: I knew him, Horatio} \\

August 3rd, 1997: the king is dead!  Long live the king.  Well, dead?  I
don't think so.  More like a minor strain.  But the media was quick to dig 
Bailey's grave for him.  The Americans once again captured the sprints.  
I was surprised the reporters didn't hail Bailey's second place finish as 
``mediocre'': he {\it only} won the silver!  The golden 9.84 was still intact, 
with Greene missing the target by 0.02s.  Was he really that far off 
the mark, though?  Table\ref{table4} reveals an awful truth.  Greene's 9.86s (+0.2 m/s) 
corrects to roughly 9.88s, which coincidentally is what a 9.84s with a 
+0.7 m/s wind becomes!  Shall we presume that the Greek god of wind was 
sitting in our corner on that fateful Sunday in Athens?

What of the other sub-9.90s performances this year, or even for all-time?
Table~\ref{table5} hints that the tail-wind has betrayed Leroy Burrell,
King Carl, Boldon, and Fredericks, whose apparent sub-9.90s were all
but same.  However, Burrell did manage to make it under that barrier,
but it wasn't with his then-WR 9.85s (+1.2 m/s) from Lausanne, nor his
9.88s (also +1.2) from the '91 WCs in Tokyo.  Rather, a seemingly
average 9.97s into a 1.3 m/s head-wind in the Barcelona '92 semi-finals
earned him that spot as a calm 9.89s (a mark which is officially ranked 
almost 80th on the all-time list!). 

And so, Greene takes the lead in the 1997 wind-corrected rankings, albeit
by a mere 0.01s over Bailey (who, in all fairness, was a mere 0.01s in
front of formidable Boldon's 9.90s in {\it actual} still-air conditions at 
Stuttgart -- no wind-correction needed!).  But all is not lost 
for the good guys.  Upon further inspection of Table~\ref{table4}, one notes that only
one name appears twice: Bailey.  While Burrell, Fredericks, and Boldon
have each legally run sub-9.90s twice, Bailey is the only one who holds that
title after all is wind-corrected! 

What a race was the 1997 Canadian Championships men's 100m final!  If we're
to believe the wind-correction figures, it propelled Bailey to be the
sole individual to hold two sub-9.90s still-air races, gave Robert Esmie
a sub-10s run to tuck under his belt, and put Canada on an even par with
the US for number of top 6 sub-10s world-class sprinters.\\

\noindent {\bf Would the {\it real} World's Fastest Man please step forward?} \\

As I write this article, there is just over a month remaining in the 1997 
Grand Prix season.  Bailey is back in force, stating in a post-relay
victory interview with NBC that his ``To do'' list for the remainder of the
season included: break World Record.  Two legal 9.91s in the same day
while injured and after coming off a virus isn't too shabby!

With Boldon and Greene also pushing their limits, and Fredericks lurking 
in the shadows, these lists may be out of date by the time they hit the 
stands.  Given the right conditions, it is very likely that we'll see the 
9.84 knocked down a couple notches.  As to how far, and more importantly, 
by whom?  Currently, almost any of these individuals can claim the title
of ``World's Fastest Man'' by some definition or another.  Although Greene 
holds the World Championship title, and in capturing gold posted the fastest 
time of the year (both wind-corrected and official), Bailey still retains 
his WR and has clocked the most sub-9.90s races ever after wind-correction.  
Meanwhile, Boldon holds the most legal sub-9.90s performances of 1997, and 
has run the fastest ever 100m/200m one-day combo (9.90/19.77 in Stuttgart).  
As for Fredericks: he's the current wind-corrected WR holder at 9.84s 
(Lausanne 1996).

Running at their current seasonal bests, a 9.83s mark can be
made by Greene with a +0.8 m/s wind, Bailey with a +1.0, and Boldon with
a +1.3 m/s tail boost.   Fredericks is currently out of the running with
his 9.97s, which would require an illegal +2.4 m/s aid.  Of course,
their current best efforts may not be their final best efforts for '97, so
anything can happen.  As of the Zurich GP (13 Aug), Bailey says he's
throwing in the towel for '97 due to ongoing injuries -- but truth and 
fiction can sometimes be hard to discern when it comes to his injuries.  
Word on the track had it that he couldn't even jog down the straight a 
week prior to the 100m final in Atlanta...
 
Which one of our contestants holds the key?  It'll be most exciting to find
out.  By the time you read this article (to paraphrase Ed McMahon): ``Someone
may already be a winner!''.

\pagebreak

\begin{table}[t]
\begin{center}
{\begin{tabular}{|l r r l|}\hline
Date & Wind-corrected & Official time & Location \\
 & time & and wind &  \\ \hline
04 May& 10.13 & (10.13, +0.0)&  Rio de Janeiro \\
08 Jun& 10.32&  (10.28, +0.6)&  Moscow  (cold rain) \\
25 Jun&  9.98 & (10.07, -1.5)&  Paris \\
02 Jul& 10.03&  (9.97, +1.0)&   Lausanne \\
19 Jul&  9.89&  (10.03, -2.1)&  Abbotsford \\
03 Aug&  9.93&  (9.91, +0.2)&   Athens WC \\ 
 & 9.94 &  (9.91, +0.5) & Athens WC (sf) \\ 
13 Aug & 10.13 & (10.17, -0.7) & Zurich \\ \hline
\end{tabular}}
\end{center}
\caption{Bailey's 1997 progression, best per meet (to 13 Aug 1997)}
\end{table}

\begin{table}[h]
\begin{center}
{\begin{tabular}{|l r r|}\hline
1. Donovan Bailey     & 9.89   &(10.03) \\
2. Robert Esmie       & 9.96   &(10.10) \\
3. Bruny Surin        & 10.01  &(10.15) \\
4. Carleton Chamers   & 10.13  &(10.27) \\
5. Glenroy Gilbert    & 10.15  &(10.29) \\
6. O'Brien Gibbons    & 10.16  &(10.30) \\
7. Troy Dos Santos    & 10.36  &(10.51) \\
8. Eric Frempong-Manso& 10.36  &(10.51) \\ \hline
\end{tabular}}
\end{center}
\caption{100m final from Abbotsford (wind -2.1 m/s), 19 Jul 1997}
\label{table2}
\end{table}
 
\begin{table}[h]
\begin{center}
{\begin{tabular}{|l l l l l|}\hline
1. Maurice Greene&9.88& (9.86, +0.2)&   03 Aug 97&      Athens WC \\
2. Donovan Bailey&  9.89 &   (10.03, -2.1)& 19 Jul 97&  Abbotsford \\
3. Ato Boldon    &  9.90 &   (9.90, +0.0)&  13 Jul 97&  Stuttgart \\
4. Tim Montgomery&  9.94 &   (9.92, +0.2)&  13 Jun 97&  Indianapolis \\
5. Frank Fredericks& 9.94 &   (9.98, -0.7)&  13 Aug 97&  Zurich \\
6. Robert Esmie  &  9.96 &   (10.10, -2.1)& 19 Jul 97&  Abbotsford \\
7. Jon Drummond  &  9.97 &   (9.92, +0.8)&  12 Jun 97&  Indianapolis \\ \hline
\end{tabular}}
\end{center}
\caption{1997 sub-10s World 100m Leaders, wind-corrected (to 13 Aug 1997); best performances per athlete}
\label{table3}
\end{table}
 
\begin{table}[h]
\begin{center}
{\begin{tabular}{|l l l l l|}\hline
9.84& Frank Fredericks & (9.86,-0.4) &   Lausanne & 03 Jul 1996\\
9.88& Maurice Greene & (9.86, +0.2) &  Athens & 03 Aug 1997\\
 & Donovan Bailey & (9.84, +0.7) &  Atlanta & 27 Jul 1996\\
9.89& Linford Christie & (9.87, +0.3) &  Stuttgart & 15 Aug 1993\\
 & Bailey  & (10.03, -2.1) & Abbotsford & 19 Jul 1997\\
 & Leroy Burrell   & (9.97, -1.3) &  Barcelona & 01 Aug 1992\\ \hline
\end{tabular}}
\end{center}
\caption{All-time wind-corrected Sub-9.90s 100m times (to 13 Aug 1997)}
\label{table4}
\end{table}

\begin{table}[h]
\begin{center}
{\begin{tabular}{|l l l l|}\hline
Rank/athlete & Official & wind & becomes \\ \hline
1. Donovan Bailey&9.84&+0.7&9.88\\
2. Leroy Burrell&9.85&+1.2&9.92\\
3. Frank Fredericks&9.86&-0.4&9.84\\
4. Maurice Greene&9.86&+0.2&9.88\\
5. Carl Lewis   &9.86&+1.2&9.93\\
6. Linford Christie&9.87&+0.3&9.89\\
7. Ato Boldon   &9.87&+1.3&9.95\\
8. Burrell      &9.88&+1.2&9.95\\
9. Frank Fredericks&9.89&+0.7&9.93\\
10. Boldon      &9.89&+0.8&9.94\\ \hline
\end{tabular}}
\end{center}
\caption{All-time official Sub-9.90s 100m times and wind-corrected equivalents (to 13 Aug 1997)}
\label{table5}
\end{table}

\begin{table}[h]
\begin{center}
{\begin{tabular}{|l l l l|}\hline
Athlete & Official & (reaction) &  Corrected \\ \hline
1. Maurice Greene (USA) & 9.86 & (+0.13) & 9.88  \\
2. Donovan Bailey (CAN)  & 9.91 & (+0.14) & 9.93 \\
3. Tim Montgomery (USA)  & 9.94 & (+0.13) & 9.96 \\
4. Frank Fredericks (NAM) & 9.95 & (+0.12) & 9.97 \\
5. Ato Boldon (TRI)      & 10.02 & (+0.12) & 10.04 \\
6. Davidson Ezinwa (NIG) & 10.10 & (+0.13) & 10.12 \\
7. Bruny Surin (CAN)     & 10.12 & (+0.14) & 10.14 \\
8. Mike Marsh (USA)      & 10.29 & (+0.14) & 10.31 \\ \hline
\end{tabular}}
\end{center}
\caption{1997 World Championships Men's 100m final; wind= +0.2 m/s}
\label{table6}
\end{table}

\end{document}